\documentclass{article}
\usepackage{amsmath}

\begin{document}

\title{Singularity Structure and Stability Analysis of the Dirac Equation on
the Boundary of the Nutku Helicoid Solution }
\author{T.Birkandan$^{\ast }$ and M. Horta\c{c}su* $^{1}$ \\
{\small $^{\ast }${\ Istanbul Technical University, Department of Physics,
Istanbul, Turkey}. }}
\maketitle

\begin{abstract}
\noindent Dirac equation written on the boundary of the Nutku helicoid space
consists of a system of ordinary differential equations. We tried to analyze
this system and we found that it has a higher singularity than those of the
Heun's equations which give the solutions of the Dirac equation in the bulk.
We also lose an independent integral of motion on the boundary. This facts
explain why we could not find the solution of the system on the boundary in
terms of known functions. We make the stability analysis of the helicoid and
catenoid cases and end up with an appendix which gives a new example where
one encounters a form of the Heun equation.
\end{abstract}

\bigskip

\bigskip


\bigskip

\bigskip \bigskip \bigskip \bigskip \bigskip \bigskip \bigskip \bigskip
\bigskip \bigskip \bigskip \bigskip \bigskip \bigskip \bigskip \bigskip
\bigskip \bigskip \bigskip \bigskip \bigskip \bigskip \footnotetext[1]{%
E-mail addresses: hortacsu@itu.edu.tr, birkandant@itu.edu.tr}

\section{Introduction}

\noindent Although one usually needs only different forms of the
hypergeometric equation or its confluent forms to describe many different
phenomena in theoretical physics, functions with higher singularity
structure are seen more and more in the literature [1-12]. A common form is
the Heun function \cite{heun}, which is studied extensively in the books by
Ronveaux and Slavyanov et al \cite{ronveaux}\cite{slavyanov}, the seminal
book by Ince \cite{Ince} as well as in several articles [17-20]. Although
for linear equations both Ince and Slavyanov et al end their singularity
analysis with Heun type functions, sometimes equations with even more
singularities are needed for relatively simple situations, which lack some
symmetries.

\noindent As an example of such a case, here we study the singularity
structure of the Dirac equations, written in the background of the Nutku
helicoid solution\cite{yavuz1}, restricted to the boundary of the helicoid.
This metric was formerly studied by Lorenz-Petzold \cite{lp}. One can study
the scalar field in this background and obtain the propagator in a closed
form \cite{yavuz2}, thanks to four integrals of motion allowed by the
metric. One can also write the Dirac equation and obtain the solutions in
terms of Mathieu functions [24-26]. One needs to study the eigenvalue
problem on the boundary to impose the boundary conditions in this problem.
The similar problem in the bulk, using partial differential equations, can
be solved in terms of known functions. On the boundary, we get a system of
ordinary differential equations. At first glance, this system seems to be
easier to analyze. When we investigate the system further, we find that this
system has higher form of singularities and fewer integrals of motion. These
turn out to be the reasons why we cannot express the solution in terms of
the functions cited in Ince's book\cite{Ince}.

\noindent In this note, we comment on the symmetries of the new problem and
study the singularity structure of the new system. We show that one gets an
equation of higher form of singularity. Since we do not have closed
solutions, we then use stability analysis to see if this system describes a
stable system. To our surprise we find that although the answer is not
affirmative for the helicoid case, we get a limit cycle for the related
catenoid solution.

\noindent Below we summarize our results. In an appendix, we show a new
example where one encounters a form of the Heun equation. This is the
solution of the laplacian in the background of the Eguchi-Hanson solution
\cite{eguchi}, trivially extended to five dimensions.

\section{Our Analysis}

\noindent The Nutku helicoid metric is given as

\begin{eqnarray}
{\normalsize ds}^{2} &=&\frac{1}{\sqrt{1+\frac{a^{2}}{r^{2}}}}%
[dr^{2}+(r^{2}+a^{2})d\theta ^{2}+\left( 1+\frac{a^{2}}{r^{2}}\sin
^{2}\theta \right) dy^{2}  \notag \\
&&-\frac{a^{2}}{r^{2}}\sin 2\theta dydz{\normalsize +}\left( 1+\frac{a^{2}}{%
r^{2}}\cos ^{2}\theta \right) {\normalsize dz}^{2}{\normalsize ]}.
\end{eqnarray}%
\noindent where $0<r<\infty $, $0\leq \theta \leq 2\pi $, $y$ and $z$ are
along the Killing directions and will be taken to be periodic coordinates on
a 2-torus \cite{yavuz2}. This is an example of a multi-center metric. This
metric reduces to the flat metric if we take $a=0$.
\begin{equation}
ds^{2}=dr^{2}+r^{2}d\theta ^{2}+dy^{2}+dz^{2}.
\end{equation}%
\noindent If we make the following transformation
\begin{equation}
r=a\sinh x,
\end{equation}%
\noindent the metric is written as
\begin{eqnarray}
ds^{2} &=&\frac{a^{2}}{2}\sinh 2x(dx^{2}+d\theta ^{2})  \notag \\
&&+\frac{2}{\sinh 2x}[(\sinh ^{2}x+\sin ^{2}\theta )dy^{2} \\
&&-\sin 2\theta dydz+(\sinh ^{2}x+\cos ^{2}\theta )dz^{2}].  \notag
\end{eqnarray}

\noindent The solutions of the Dirac equation, written in the background of
the Nutku helicoid metric, can be expressed as a special form of Heun
functions \cite{tolga1}\cite{tolga2}. We could reduce the double confluent
Heun function obtained for the radial equation to the Mathieu function with
coordinate transformations. Mathieu function is a related but much more
studied function with similar singularity structure. In the work cited
above, \cite{tolga1}\cite{tolga2}, we tried to get the solution of the
little Dirac equation, the name used for the equation restricted to a
boundary of the helicoid. We needed this solution also to be able to
calculate the index of the differential operator. For a fixed value of the
radial coordinate we had only a coupled system of ordinary differential
equations, which, in general, should be much simpler to solve than the
coupled system of partial differential equations obtained for the full Dirac
equation, written for the bulk. We were successful to obtain the solution in
this latter case in terms of Mathieu functions. We were not able to identify
the solutions for the little Dirac equation , though.

\noindent The first thing we check is whether we lose any of the three
Killing vectors and one Killing tensor. For the metric in question, one has
three integrals of motion, \cite{yavuz2} namely $p_{y},p_{z}$ and $g^{\mu
\nu }p_{\mu }p_{\nu }={\mu^{2}}$, namely
\begin{equation}
\left( \frac{dS_{x}}{dx}\right) ^{2}+\left( \frac{dS_{\theta }}{d\theta }%
\right) ^{2}+a^{2}\,\left( p_{y}^{2}+p_{z}^{2}\right) \,\sinh ^{2}x-\frac{%
\mu ^{2}\,a^{2}}{2}\,\sinh 2x+\,a^{2}\,\left( \cos \theta \,p_{y}\,+\,\sin
\theta p_{z}\right) ^{2}=0,
\end{equation}
\noindent and an extra integral of motion, the Killing tensor \cite{yavuz2}
,
\begin{equation}
K=-p_{\theta }^{2}-a^{2}(cos\theta p_{y}+sin\theta p_{z})^{2}
\end{equation}
which gives us the angular equation for a fixed value of the constant $%
\lambda $ . See eq.s (46) and (47) of \cite{yavuz2}. When one restricts the
solution to a fixed value of the radial coordinate, the value of $g^{\mu \nu
}p_{\mu }p_{\nu }$ is not an independent constant of motion from the other
two Killing vectors and the Killing tensor.

\noindent Trying to see the effect of this result on our problem, we
investigate the singularity structure of the equation we get for the little
Dirac operator. Here we study the simplest case, where the eigenvalue $%
\lambda $ is equal to zero, since the answer to our problem is already
apparent here. For this case, instead of getting a system of four coupled
equations we get coupling only between two of them at a time. When we
analyze the system by reducing them to a second order equation for a single
dependent variable, we find an operator with two irregular and one regular
singularities, which is one more than allowed for the equations considered
among the Heun functions. The double confluent Heun function, the solution
one obtains for the full Dirac equation, has two irregular singularities,
missing the extra regular singularity of the case studied here.

\noindent To make our discussion concrete we explicitly perform the
calculation in the next section.

\subsection{Singularities}

\noindent The Dirac equation written in the background of the Nutku
helicoids metric is written as
\begin{equation}
(\partial _{x}+i\partial _{\theta })\Psi _{3}\ +iak[cos(\theta -\phi
+ix)]\Psi _{4}=0,
\end{equation}%
\begin{equation}
(\partial _{x}-i\partial _{\theta })\Psi _{4}\ -iak[cos(\theta -\phi
-ix)]\Psi _{3}=0,
\end{equation}%
\begin{equation}
(-\partial _{x}+i\partial _{\theta })f_{1}\ +iak[cos(\theta -\phi +ix)]f_{2}{%
=0},
\end{equation}%
\begin{equation}
(-\partial _{x}-i\partial _{\theta })f_{2}\ -iak[cos(\theta -\phi
-ix)]f_{1}=0.
\end{equation}

\noindent These equations have simple solutions \cite{Nuri} which can also
be expanded in terms of products of radial and angular Mathieu functions
\cite{Chaos}\cite{tolga1}. Problem arises when these solutions are
restricted to boundary \cite{tolga2}.

\noindent To impose these boundary conditions we need to write the little
Dirac equation, the Dirac equation restricted to the boundary, where the
variable $x$ takes a fixed value $x_{0}$. We choose to write the equations
in the form,
\begin{equation}
{\frac{\sqrt{2}}{{a}}}\{i\frac{d}{d\theta }\Psi _{3}\ +ikacos(\theta -\phi
+ix_{0})\Psi _{4}\}=\lambda f_{1},
\end{equation}%
\begin{equation}
{\frac{\sqrt{2}}{{a}}}\{-i\frac{d}{d\theta }\Psi _{4}\ -iakcos(\theta -\phi
-ix_{0})\Psi _{3}\}=\lambda f_{2},
\end{equation}%
\begin{equation}
{\frac{\sqrt{2}}{{a}}}\{-i\frac{d}{d\theta }f_{1}\ -iakcos(\theta -\phi
+ix_{0})f_{2}\}=\lambda \Psi _{3},
\end{equation}%
\begin{equation}
{\frac{\sqrt{2}}{{a}}}\{i\frac{d}{d\theta }f_{2}\ +iakcos(\theta -\phi
-ix_{0})f_{1}\}=\lambda \Psi _{4}.
\end{equation}%
\noindent Here $\lambda $ is the eigenvalue of the little Dirac equation. We
take $\lambda =0$ as the simplest case. The transformation

\begin{equation}
\Theta =\theta -\phi -ix_{0}
\end{equation}

\noindent can be used. Then we solve $f_{1\text{ }}$ in the latter two
equations in terms of $f_{2}$:%
\begin{equation}
-\frac{d^{2}}{d\Theta ^{2}}f_{2}-\tan \Theta \frac{d}{d\Theta }f_{2}+\frac{%
(ak)^{2}}{2}[\cos (2\Theta )\cosh (2x_{0})-i\sin (2\Theta )\sinh
(2x_{0})+\cosh (2x_{0})]f_{2}=0
\end{equation}

\noindent When we make the transformation%
\begin{equation}
u=e^{2i\Theta },
\end{equation}

\noindent the equation reads,%
\begin{equation}
\{4(u+1)u[u\frac{d^{2}}{du^{2}}+\frac{d}{du}]-2iu(u-1)\frac{d}{du}+\frac{%
(ak)^{2}}{2}(u+1)[ue^{-2x_{0}}+\frac{1}{u}e^{2x_{0}}+\cosh
(2x_{0})]\}f_{2}=0.  \label{denklem}
\end{equation}

\noindent This equation has irregular singularities at $u=0$ and $\infty $
and a regular singularity at $u=-1$. \noindent If we try a solution in the
form $\overset{\infty }{\underset{n=-\infty }{\sum }}a_{n}u^{n}$ \noindent
around the irregular singularity $u=0$ we end up with a four-term recursion
relation as

\begin{eqnarray}
&&a_{n-1}[4(n^{2}-2n+1)-2i(n-1)+\frac{(ak)^{2}}{2}(\frac{3}{2}e^{-2x_{0}}+%
\frac{1}{2}e^{2x_{0}})]  \notag \\
&&+a_{n}[4n^{2}+2in+\frac{(ak)^{2}}{2}(\frac{3}{2}e^{-2x_{0}}+\frac{1}{2}%
e^{2x_{0}})]+ \\
&&a_{n-2}[\frac{(ak)^{2}}{2}e^{-2x_{0}}]+a_{n+1}[\frac{(ak)^{2}}{2}%
e^{2x_{0}}]  \notag
\end{eqnarray}

\noindent As it is known, in the Heun equation case, this kind of series
solution gives a three-term relation \cite{r155}.

\noindent If we search for a solution of the Thom\'{e} type we may try a
solution of the form $f_{2}=e^{\frac{A}{\sqrt{u}}}g(u)$. This form does not
allow us to get a Taylor series expansion around the irregular point $u=0$
\cite{slav115} \cite{olver}.

\noindent If we try a series solution around the regular singularity at $%
u=-1 $ as

\noindent $\overset{\infty }{\underset{n=0}{\sum }}a_{n}(u+1)^{n+\alpha }$
we find a relation between five consecutive coefficients for the solution.
Therefore, we may conclude that the solution of this equation cannot be
written in terms of Heun functions or simplier special functions.

\noindent To check this further, we first set the coefficient of $\frac{1}{u}
$ term in equation \ref{denklem} equal to zero to change our irregular
singularity at zero to a regular one. Then we keep this term and discard the
$ue^{-2x_{0}}$ term to reduce the singularity structure of infinity. In both
cases one can check that the solution can be expressed in terms of confluent
Heun functions. This shows that reducing one of the singularities yields a
Heun function. Thus, we conclude that the full equation \ref{denklem} is not
one of the better known equations in the literature, which are included in
the computer packages like Maple, cited in the seminal book by Ince \cite%
{Ince}.

\noindent To investigate the type of our equation we try to get a confluent
form of a new equation,%
\begin{equation}
y^{\prime \prime }(z)+(\frac{1-\mu _{0}}{z}+\frac{1-\mu _{1}}{z+1}+\frac{%
1-\mu _{2}}{z-a})y^{\prime }(z)+\frac{\beta _{0}+\beta _{1}z+\beta _{2}z^{2}%
}{z^{2}(z-a)}y(z)=0  \label{bizim}
\end{equation}

\noindent with regular singularities at $0$, $-1$ and $a$ and an irregular
singularity at infinity. This equation differs from the generalized Heun
equation \cite{schmid}\cite{schafke}:%
\begin{equation}
y^{\prime \prime }(z)+(\frac{1-\mu _{0}}{z}+\frac{1-\mu _{1}}{z+1}+\frac{%
1-\mu _{2}}{z-a}-\alpha )y^{\prime }(z)+\frac{\beta _{0}+\beta _{1}z+\beta
_{2}z^{2}}{z(z+1)(z-a)}y(z)=0  \label{genheun1}
\end{equation}

\noindent which also has regular singularities at $0$, $-1$ and $a$\ and an
irregular singularity at infinity. These two equations both have four-term
recursion relations. They, however, have different singularity ranks
according to the classification given in \cite{ronveaux293}. When we put $a=0
$ in equation \ref{bizim}, we get%
\begin{equation}
y^{\prime \prime }(z)+(\frac{2-\mu _{0}-\mu _{2}}{z}+\frac{1-\mu _{1}}{z+1}%
)y^{\prime }(z)+\frac{\beta _{0}+\beta _{1}z+\beta _{2}z^{2}}{z^{3}}y(z)=0.
\label{baskabizim}
\end{equation}

\noindent We get a singularity structure as a regular singularity at $-1$
and two irregular singularities at zero and infinity like the equation \ref%
{denklem}.

\noindent Both equations \ref{denklem} and \ref{baskabizim} have four-term
recursion relations when a Laurent power series solution is attempted. We
may name this equation as the confluent form of the equation \ref{bizim}. It
is in the same form as our original equation rewritten as
\begin{equation}
\{\frac{d^{2}}{du^{2}}+[\frac{1}{u}(1+\frac{i}{2})+\frac{i}{u+1}]\frac{d}{du}%
+\frac{(ak)^{2}}{2}[\frac{e^{-2x_{0}}}{u}+\frac{e^{2x_{0}}}{u^{3}}+\frac{%
\cosh (2x_{0})}{u^{2}}]\}f_{2}=0.
\end{equation}

\noindent Both of these equations have s-rank multisymbols $\{1,\frac{3}{2},%
\frac{3}{2}\}$ referring to the singularities at $\{-1,0,\infty \}$ \cite%
{ronveaux293}.

\noindent We could not obtain a confluent equation similar to the equation %
\ref{denklem} from the generalized Heun equation \ref{genheun1}. If we
simply put $a=0$ in this equation we get the confluent Heun solution. We can
obtain an equation with the same singularity structure as our equation only
if we write the equation,%
\begin{equation}
y^{\prime \prime }(z)+(\frac{1-\mu _{0}}{z}+\frac{1-\mu _{1}}{z+1}+\frac{%
1-\mu _{2}}{z-a})y^{\prime }(z)+\frac{\frac{\beta _{-1}}{z}+\beta _{0}+\beta
_{1}z+\beta _{2}z^{2}}{z(z+1)(z-a)}y(z)=0,  \label{beta3}
\end{equation}

\noindent and make $a$ approach zero. Then we end up with an equation having
the same singularity structure as in equation \ref{denklem}.

\noindent If we want to compare equation \ref{denklem} with equation \ref%
{genheun1} we have to form a confluent form of the latter equation. To
coalesce the singularities at zero, we make a detour and then use standart
techiques \cite{filip}. We first translate the singularity at zero to a
singularity at $b$,%
\begin{equation}
y^{\prime \prime }(z)+(\frac{1-\mu _{0}}{z-b}+\frac{1-\mu _{1}}{z+1}+\frac{%
1-\mu _{2}}{z-a}-\alpha )y^{\prime }(z)+\frac{\beta _{0}+\beta _{1}z+\beta
_{2}z^{2}}{(z-b)(z+1)(z-a)}y(z)=0  \label{genheun}
\end{equation}

\noindent This equation has regular singularities at $z=-1,a,b$ and an
irregular singularity at infinity. We make the transformation $z=\frac{1}{v}$%
. Then, the equation \ref{genheun} becomes,%
\begin{eqnarray}
&&y^{\prime \prime }(v)+(\frac{2-(1-\mu _{0})-(1-\mu _{1})-(1-\mu _{2})}{v}+%
\frac{\alpha }{v^{2}}  \notag \\
&&-\frac{1-\mu _{0}}{\frac{1}{b}-v}-\frac{1-\mu _{1}}{v+1}-\frac{1-\mu _{2}}{%
\frac{1}{a}-v})y^{\prime }(v)  \notag \\
&&+\frac{\frac{\beta _{0}}{v}+\frac{\beta _{1}}{v^{2}}+\frac{\beta _{2}}{%
v^{3}}}{(1-bv)(1+v)(1-av)}y(v)=0
\end{eqnarray}

\noindent We set $\mu _{0}=-\mu _{1}=1/b=2/a=\epsilon $ and take the limit $%
\epsilon \rightarrow \infty .$ Then we transform back to the original
variables using $v=\frac{1}{z}$ to obtain%
\begin{equation}
y^{\prime \prime }(z)+(\frac{3-\mu _{1}}{z}-\alpha -\frac{\mu _{1}-1}{z(z+1)}%
+\frac{1}{z^{2}})y^{\prime }(z)+\frac{\beta _{0}+\beta _{1}z+\beta _{2}z^{2}%
}{z^{2}(z+1)}y(z)=0  \label{onda}
\end{equation}

\noindent This equation is a confluent form of the equation \ref{genheun}.
It has a regular singularity at $-1$ and two irregular singularities at zero
and infinity and a four-way recursion relation when expanded around zero
using a Laurent expansion. This equation has s-rank multisymbols $\{1,2,2\}$
referring to singularities at $\{-1,0,\infty \}$ \cite{ronveaux293}. Here we
get the rank-2 irregular singularities at zero and infinity only from the
coefficient of the first derivative whereas in equation \ref{denklem} and in
equation \ref{baskabizim} the coefficient of the term without derivatives
gives us these singularities. They also have different ranks. Even if we set
$\alpha =0$ in equation \ref{onda} then we get rank=$\{1,2,\frac{3}{2}\}$
which is different from the rank of our original equation. Hence, our
equation \ref{denklem} may be a confluent form only of a variation of the
equation \ref{genheun1}, like equation \ref{beta3}.

\subsection{Stability analysis for the little Dirac Equation}

\noindent The little Dirac equation is a system of linear differential
equations with periodic coefficients. Then we can write the system as \cite%
{bellman}\cite{codlev}:

\begin{equation}
\partial _{\theta }\psi =P(\theta )\psi
\end{equation}

\noindent Here $P(\theta +\tau )=P(\theta )$ and $\tau $ is the period of
the coefficients. $(\tau \neq 0)$. According to Bellman, if $\psi (0)=I$ ,
we can write
\begin{equation}
\psi (\theta )=Q(\theta )e^{B\theta. }
\end{equation}

\noindent The matrix $Q(\theta )$ is also periodic with the period $\tau $.
Then we have%
\begin{eqnarray}
\psi (\theta +\tau ) &=&Q(\theta +\tau )e^{B\theta }e^{B\tau }  \notag \\
&=&Q(\theta )e^{B\theta }e^{B\tau } \\
&=&\psi (\theta )e^{B\tau }.  \notag
\end{eqnarray}

\noindent We can define $e^{B\tau }\equiv C$ and use

\begin{equation}
C=\psi (\theta )^{-1}\psi (\theta +\tau )
\end{equation}

\noindent to obtain $C$. The Jordan normal form of $C$, $T$ being the
transformation matrix,

\begin{equation}
C=T%
\begin{pmatrix}
L_{1} &  &  \\
& \ddots &  \\
&  & L_{r}%
\end{pmatrix}%
T^{-1}
\end{equation}

\noindent gives us the $B$ matrix:

\begin{equation}
B=\frac{1}{\tau }\ln L
\end{equation}

\noindent The eigenvalues of $B$ give us the characteristic roots. We will
use these characteristic roots with different parameters in our stability
analysis. The characteristic roots are given by $\alpha _{i}$, ($i=1..4$).

\begin{center}
Table 1. The change in the characteristic roots with respect to parameters
(helicoid case)

\begin{tabular}{|c|c|c|c|c|c|}
\hline
$a$ & $k$ & $x_{0}$ & $\lambda $ &
\begin{tabular}{l}
real parts of the \\
characteristic roots ($\times 2\pi $)%
\end{tabular}
&
\begin{tabular}{l}
signs of the \\
characteristic roots%
\end{tabular}
\\ \hline
$1$ & $1$ & $1$ & $1$ & $\alpha _{1,2,3,4}=9.25029$ & $++--$ \\ \hline
$0.5$ & $"$ & $"$ & $"$ & $\alpha _{1,2,3,4}=4.03926$ & $"$ \\ \hline
$0.8$ & $"$ & $"$ & $"$ & $\alpha _{1,2,3,4}=7.2163$ & $"$ \\ \hline
$1.1$ & $"$ & $"$ & $"$ & $\alpha _{1,2,3,4}=10.2542$ & $"$ \\ \hline
$1.5$ & $"$ & $"$ & $"$ & $\alpha _{1,2,3,4}=14.2215$ & $"$ \\ \hline
$1$ & $0.5$ & $1$ & $1$ & $\alpha _{1,2,3,4}=5.8327$ & $"$ \\ \hline
$"$ & $0.8$ & $"$ & $"$ & $\alpha _{1,2,3,4}=7.75421$ & $"$ \\ \hline
$"$ & $1.1$ & $"$ & $"$ & $\alpha _{1,2,3,4}=10.0319$ & $"$ \\ \hline
$"$ & $1.5$ & $"$ & $"$ & $\alpha _{1,2,3,4}=13.2745$ & $"$ \\ \hline
$1$ & $1$ & $0.5$ & $1$ & $\alpha _{1,2,3,4}=6.56206$ & $"$ \\ \hline
$"$ & $"$ & $0.8$ & $"$ & $\alpha _{1,2,3,4}=7.91586$ & $"$ \\ \hline
$"$ & $"$ & $1.1$ & $"$ & $\alpha _{1,2,3,4}=10.0632$ & $"$ \\ \hline
$"$ & $"$ & $1.5$ & $"$ & $\alpha _{1,2,3,4}=14.4648$ & $"$ \\ \hline
$1$ & $1$ & $1$ & $0.5$ & $\alpha _{1,2,3,4}=8.30164$ & $"$ \\ \hline
$"$ & $"$ & $"$ & $0.8$ & $\alpha _{1,2,3,4}=8.81342$ & $"$ \\ \hline
$"$ & $"$ & $"$ & $1.1$ & $\alpha _{1,2,3,4}=9.49249$ & $"$ \\ \hline
$"$ & $"$ & $"$ & $1.5$ & $\alpha _{1,2,3,4}=10.5888$ & $"$ \\ \hline
\end{tabular}
\end{center}

\noindent \.Our calculations indicate that $f_{1}$ and $f_{2}$ solutions are
not stable (positive characteristic root) while, $\Psi _{3}$ and $\Psi _{4} $
solutions are stable (negative characteristic root). As it is seen in Table
1, when we keep all the other parameters constant and vary only $a$, the
value of the roots are influenced most, whereas the effect of the variation
in the value of $\lambda $ changes the value of the roots least. We also
find that when these parameters exceed unity in absolute value, we encounter
inconsistencies in the numerical values. The separation between consecutive
roots increase and some negative roots go to positive values for large
values of the parameters. If we keep the values of the parameters in the
range $[-1,1] $, we seem to have no such problems.

\noindent The periodicity of the defined $Q$ can be checked using numerical
means.

\noindent We use
\begin{equation}
\psi (\theta +\tau )=Q(\theta +\tau )e^{B\theta }C \\
\end{equation}

\noindent and for $\theta =0$, $\psi (\theta )=Q(\theta )e^{B\theta }$ to
give,

\begin{equation}
\psi (\tau )C^{-1}=Q(\tau )=Q(0)=\psi (0)=I .
\end{equation}

\noindent We check numerically that this equation is satisfied; hence, $Q$
is periodic.

\noindent

\noindent For the catenoid case, one replaces $a$ with $ia$ in the metric.
The same stability procedure is performed for this case and we find the
characteristic roots given in the Table 2.

\begin{center}
Table 2. The change in the characteristic roots with respect to parameters
(catenoid case)

\begin{tabular}{|c|c|c|c|c|c|}
\hline
$a$ ($\times $ $i$) & $k$ & $x_{0}$ & $\lambda $ & characteristic roots ($%
\times 2\pi $) &
\begin{tabular}{l}
signs of the \\
characteristic roots%
\end{tabular}
\\ \hline
$1$ & $1$ & $1$ & $1$ & $%
\begin{tabular}[b]{l}
$\alpha _{1,2}=-5.84543\times 10^{-8}+2.08426i$ \\
$\alpha _{3,4}=-6.40967\times 10^{-8}+0.72266i$%
\end{tabular}%
$ & $+-+-$ \\ \hline
$0.5$ & $"$ & $"$ & $"$ & $%
\begin{tabular}[b]{l}
$\alpha _{1,2}=-2.40553\times 10^{-7}-1.19724i$ \\
$\alpha _{3,4}=-2.49225\times 10^{-7}-2.3721i$%
\end{tabular}%
$ & $-+-+$ \\ \hline
$0.8$ & $"$ & $"$ & $"$ & $%
\begin{tabular}[b]{l}
$\alpha _{1,2}=-5.63046\times 10^{-7}-0.233835i$ \\
$\alpha _{3,4}=-5.78178\times 10^{-7}-2.42696i$%
\end{tabular}%
$ & $-+-+$ \\ \hline
$1.1$ & $"$ & $"$ & $"$ & $%
\begin{tabular}[b]{l}
$\alpha _{1,2}=-7.82856\times 10^{-7}+2.9982i$ \\
$\alpha _{3,4}=-7.88358\times 10^{-7}-0.186969i$%
\end{tabular}%
$ & $+-+-$ \\ \hline
$1.5$ & $"$ & $"$ & $"$ & $%
\begin{tabular}[b]{l}
$\alpha _{1,2}=-7.53539\times 10^{-7}-0.526163i$ \\
$\alpha _{3,4}=-7.7255\times 10^{-7}-2.35978i$%
\end{tabular}%
$ & $-+-+$ \\ \hline
$1$ & $0.5$ & $1$ & $1$ & $%
\begin{tabular}[b]{l}
$\alpha _{1,2}=3.49081\times 10^{-8}+0.661356i$ \\
$\alpha _{3,4}=2.02794\times 10^{-8}-1.31726i$%
\end{tabular}%
$ & $+--+$ \\ \hline
$"$ & $0.8$ & $"$ & $"$ & $%
\begin{tabular}[b]{l}
$\alpha _{1,2}=-4.04957\times 10^{-7}+0.864554i$ \\
$\alpha _{3,4}=-4.22823\times 10^{-7}+2.3638i$%
\end{tabular}%
$ & $+-+-$ \\ \hline
$"$ & $1.1$ & $"$ & $"$ & $%
\begin{tabular}[b]{l}
$\alpha _{1,2}=-2.2577\times 10^{-7}+2.7328i$ \\
$\alpha _{3,4}=-2.30929\times 10^{-7}+0.120058i$%
\end{tabular}%
$ & $+-+-$ \\ \hline
$"$ & $1.5$ & $"$ & $"$ & $%
\begin{tabular}[b]{l}
$\alpha _{1,2}=-1.33916\times 10^{-7}+0.715427i$ \\
$\alpha _{3,4}=-1.38904\times 10^{-7}-2.75007i$%
\end{tabular}%
$ & $+--+$ \\ \hline
$1$ & $1$ & $0.5$ & $1$ & $%
\begin{tabular}[b]{l}
$\alpha _{1,2}=-4.68625\times 10^{-8}-0.232525i$ \\
$\alpha _{3,4}=-5.34771\times 10^{-8}-2.02431i$%
\end{tabular}%
$ & $-+-+$ \\ \hline
$"$ & $"$ & $0.8$ & $"$ & $%
\begin{tabular}[b]{l}
$\alpha _{1,2}=-4.69901\times 10^{-7}-1.06977i$ \\
$\alpha _{3,4}=-4.72986\times 10^{-7}-2.11709i$%
\end{tabular}%
$ & $-+-+$ \\ \hline
$"$ & $"$ & $1.1$ & $"$ & $%
\begin{tabular}[b]{l}
$\alpha _{1,2}=-6.06425\times 10^{-7}-2.74153i$ \\
$\alpha _{3,4}=-6.14188\times 10^{-7}-0.143986i$%
\end{tabular}%
$ & $-+-+$ \\ \hline
$"$ & $"$ & $1.5$ & $"$ & $%
\begin{tabular}[b]{l}
$\alpha _{1,2}=-4.26375\times 10^{-7}+0.299098i$ \\
$\alpha _{3,4}=-4.4214\times 10^{-7}+1.55822i$%
\end{tabular}%
$ & $+-+-$ \\ \hline
$1$ & $1$ & $1$ & $0.5$ & $%
\begin{tabular}[b]{l}
$\alpha _{1,2}=-4.67773\times 10^{-7}+0.761149i$ \\
$\alpha _{3,4}=-4.69627\times 10^{-7}+0.741915i$%
\end{tabular}%
$ & $+-+-$ \\ \hline
$"$ & $"$ & $"$ & $0.8$ & $%
\begin{tabular}[b]{l}
$\alpha _{1,2}=-2.82256\times 10^{-7}-1.5208i$ \\
$\alpha _{3,4}=-2.86571\times 10^{-7}-0.79892i$%
\end{tabular}%
$ & $-+-+$ \\ \hline
$"$ & $"$ & $"$ & $1.1$ & $%
\begin{tabular}[b]{l}
$\alpha _{1,2}=-4.14479\times 10^{-7}+2.38047i$ \\
$\alpha _{3,4}=-4.1985\times 10^{-7}+0.650337i$%
\end{tabular}%
$ & $+-+-$ \\ \hline
$"$ & $"$ & $"$ & $1.5$ & $%
\begin{tabular}[b]{l}
$\alpha _{1,2}=-3.33914\times 10^{-7}-2.62503i$ \\
$\alpha _{3,4}=-3.42067\times 10^{-7}-0.145628i$%
\end{tabular}%
$ & $-+-+$ \\ \hline
\end{tabular}%
\\[0pt]

\bigskip
\end{center}

\noindent We see that the real parts of these roots are compatible with
assigning to value zero within numerical errors. This corresponds to a limit
cycle$\cite{hurewicz}\cite{lefschetz}$.

\noindent

\section{Conclusion}

\noindent Here we performed a systematic analysis of the Dirac equation
restricted to the boundary when it is written in the background of the Nutku
helicoid solution \cite{yavuz2} We find that the resulting system of
ordinary differential equations has a singularity which is higher than those
of the Heun functions which are solutions for the bulk. We also lose an
independent integral of motion. This fact explains why we could not obtain
the solution of the system on the boundary in terms of well known functions.

\noindent The stability analysis we performed shows that although this
system is not stable, a related system, the catenoid solution is. We can,
thus, give a meaning to its solutions, although we can not get explicit
solutions for the little Dirac equation obtained from it too. \bigskip

\section{Appendix: Scalar field in the background of the extended
Eguchi-Hanson solution}

\noindent\ To go to five dimensions, we can add a time component to the
Eguchi-Hanson metric \cite{eguchi}\ so that we have
\begin{equation}
ds^{2}=-dt^{2}+{\frac{{1}}{{1-{\frac{{a^{4}}}{{r^{4}}}}}}}%
dr^{2}+r^{2}(\sigma _{x}^{2}+\sigma _{y}^{2})+r^{2}(1-{\frac{{a^{4}}}{{r^{4}}%
}})\sigma _{z}^{2}
\end{equation}%
\noindent where
\begin{equation}
\sigma _{x}={\frac{{1}}{{2}}}(-\cos \xi d\theta -\sin \theta \sin \xi d\phi )
\end{equation}%
\begin{equation}
\sigma _{y}={\frac{{1}}{{2}}}(\sin \xi d\theta -\sin \theta \cos \xi d\phi )
\end{equation}%
\begin{equation}
\sigma _{z}={\frac{{1}}{{2}}}(-d\xi -\cos \theta d\phi ).~
\end{equation}%
\noindent This is a vacuum solution.

\noindent If we take
\begin{equation}
\Phi =e^{ikt}e^{in\phi }e^{i(m+{\frac{{1}}{{2}}})\xi }\varphi (r,\theta ),
\end{equation}%
\noindent we find the scalar equation as

\begin{eqnarray}
H\varphi (r,\theta ) &=&({\frac{{r^{4}-a^{4}}}{{r^{2}}}}\partial _{rr}+{%
\frac{{3r^{4}+a^{4}}}{{r^{3}}}}\partial _{r}+k^{2}r^{2}+{\frac{{4a^{4}m^{2}}%
}{{a^{4}-r^{4}}}}+  \notag \\
&&4\partial _{\theta \theta }+4\cot \theta \partial _{\theta }+{\frac{{%
8mn\cos \theta -4(m^{2}+n^{2})}}{{\sin ^{2}\theta }})}\varphi (r,\theta ).
\end{eqnarray}

\noindent If we take $\varphi (r,\theta )=f(r)g(\theta )$, the solution of
the radial part is expressed in terms of confluent Heun ($\mathit{H}_{C}$)
functions.

\begin{equation*}
f\left( r\right) =\left( -a^{4}+r^{4}\right) ^{{\frac{{1}}{{2}}}\,m}\
\mathit{H}_{C}\left( 0,m,m,{\frac{{1}}{{2}}}\,{k}^{2}{a}^{2},{\frac{{1}}{{2}}%
}\,{m}^{2}-{\frac{{1}}{{4}}}\,\lambda -{\frac{{1}}{{4}}}\,{k}^{2}{a}^{2},\,{{%
\frac{{{a}^{2}+{r}^{2}}}{{2{a}^{2}}}}}\right) \
\end{equation*}%
\begin{equation}
+\left( {a}^{2}+{r}^{2}\right) ^{-{\frac{{1}}{{2}}}\,m}\left(
r^{2}-a^{2}\right) ^{{\frac{{1}}{{2}}}\,m}\mathit{H}_{C}\left( 0,-m,m,{\frac{%
{1}}{{2}}}\,{k}^{2}{a}^{2},{\frac{{1}}{{2}}}\,{m}^{2}-{\frac{{1}}{{4}}}%
\,\lambda -{\frac{{1}}{{4}}}\,{k}^{2}{a}^{2},{{\frac{{{a}^{2}+{r}^{2}}}{{2{a}%
^{2}}}}}\right)
\end{equation}%
\qquad

\noindent The angular solution is in terms of hypergeometric solutions.

\begin{equation*}
g\left( \theta \right) ={\frac{{1}}{{\sin \theta }}}{\{\sqrt{2-2\,\cos
\left( \theta \right) }\left( {\frac{{1}}{{2}}}\,\cos \left( \theta \right) -%
{\frac{{1}}{{2}}}\right) ^{{\frac{{1}}{{2}}}\,m}\left( {\frac{{1}}{{2}}}%
\,\cos \left( \theta \right) -{\frac{{1}}{{2}}}\right) ^{-{\frac{{1}}{{2}}}%
\,n}}\
\end{equation*}%
\begin{equation*}
{[\left( 2\,\cos \left( \theta \right) +2\right) ^{{\frac{{1}}{{2}}}-{\frac{{%
1}}{{2}}}\,n-{\frac{{1}}{{2}}}\,m}}\
\end{equation*}%
\begin{equation}
\times {\mathit{_{2}F_{1}}}({[-n+{\frac{{1}}{{2}}}\,\sqrt{\lambda +1}+{\frac{%
{1}}{{2}}},}\ {-n-{\frac{{1}}{{2}}}\,\sqrt{\lambda +1}+{\frac{{1}}{{2}}}%
],[1-n-m],{\frac{{1}}{{2}}}\,\cos \left( \theta \right) +{\frac{{1}}{{2}}})}%
\
\end{equation}%
\begin{equation*}
{+\left( 2\,\cos \left( \theta \right) +2\right) ^{{\frac{{1}}{{2}}}+{\frac{{%
1}}{{2}}}\,n+{\frac{{1}}{{2}}}\,m}}\
\end{equation*}%
\begin{equation*}
\times {\mathit{_{2}F_{1}}([m+{\frac{{1}}{{2}}}\,\sqrt{\lambda +1}+{\frac{{1}%
}{{2}}},m-{\frac{{1}}{{2}}}\,\sqrt{\lambda +1}+{\frac{{1}}{{2}}}],}\ {%
[1+n+m],{\frac{{1}}{{2}}}\,\cos \left( \theta \right) +{\frac{{1}}{{2}}})]\}}
\end{equation*}

\noindent If the variable transformation $r=a\sqrt{\cosh x}$ is made, the
solution can be expressed as

\begin{equation*}
f\left( x\right) ={\frac{{1}}{{\sinh x}}}{\{\left( \sinh \left( x\right)
\right) ^{m+1}}\ \mathit{H}_{C}{\left( 0,m,m,{\frac{{1}}{{2}}}\,{k}^{2}{a}%
^{2},{\frac{{1}}{{2}}}\,{m}^{2}-{\frac{{1}}{{4}}}\,\lambda -{\frac{{1}}{{4}}}%
\,{k}^{2}{a}^{2},{\frac{{1}}{{2}}}\,\cosh \left( x\right) +{\frac{{1}}{{2}}}%
\right) }\
\end{equation*}%
\begin{equation}
{+\left( 2\,\cosh \left( x\right) +2\right) ^{-{\frac{{1}}{{2}}}\,m+{\frac{{1%
}}{{2}}}}\left( 2\,\cosh \left( x\right) -2\right) ^{{\frac{{1}}{{2}}}\,m+{%
\frac{{1}}{{2}}}}}
\end{equation}%
\begin{equation*}
\times \mathit{H}_{C}{\left( 0,-m,m,{\frac{{1}}{{2}}}\,{k}^{2}{a}^{2},{\frac{%
{1}}{{2}}}\,{m}^{2}-{\frac{{1}}{{4}}}\,\lambda -{\frac{{1}}{{4}}}\,{k}^{2}{a}%
^{2},{\frac{{1}}{{2}}}\,\cosh \left( x\right) +{\frac{{1}}{{2}}}\right) \}}.
\end{equation*}%
\noindent We tried to express the equation for the radial part in terms of $%
u={\frac{{a^{2}+r^{2}}}{{2a^{2}}}}$ to see the singularity structure more
clearly. Then the radial differential operator reads
\begin{equation}
4{\frac{{d^{2}}}{{du^{2}}}}+4\left( {\frac{{1}}{{u-1}}}+{\frac{{1}}{{u}}}%
\right) {\frac{{d}}{{du}}}+k^{2}a^{2}\left( {\frac{{1}}{{u-1}}}+{\frac{{1}}{{%
u}}}\right) +{\frac{{m^{2}}}{{u^{2}(1-u)^{2}}}}.
\end{equation}

\noindent This operator has two regular singularities at zero and one, and
an irregular singularity at infinity, the singularity structure of the
confluent Heun equation. This is different from the hypergeometric equation,
which has regular singularities at zero, one and infinity.

\noindent \textbf{Acknowledgement}: We would like to thank Prof. Ay\c{s}e
Bilge for correspondence and discussions. This work is supported by T\"{U}B%
\.{I}TAK, the Scientific and Technological Council of Turkey. The work of
M.H. is also supported by T\"{U}BA, the Academy of Sciences of Turkey.

\end{document}